\documentclass[12pt,a4paper,twoside]{article}
\def\runninghead#1#2{\pagestyle{myheadings}
\markboth{{\protect\footnotesize\it{\quad #2}}\hfill}
{\hfill{\protect\footnotesize\it{#2\quad}}}} \headsep=15pt
\newcommand{\be}{\begin{equation}}
\newcommand{\ee}{\end{equation}}
\newcommand{\bea}{\begin{eqnarray}}
\newcommand{\eea}{\end{eqnarray}} %
 \title{Torsion, Chern-Simons Term and Diffeomorphism Invariance}
 \author{Prasanta Mahato\footnote{Narasinha Dutt College,
         Howrah, West Bengal, India 711101, e-mail: pmahato@dataone.in}\\Partha Bhattacharya \footnote{Bangabasi Morning College, Kolkata, West Bengal, India 700009, e-mail: parthab68@yahoo.co.in}}
\date{}
\begin{document}
\runninghead{\underline{Prasanta Mahato}}{\underline{Torsion, Chern-Simons Term and ....}}
\maketitle
\setcounter{footnote}{0}
 \begin{abstract}
 In the $torsion \otimes curvature$ approach of gravity Chern-Simons modification  has been considered here. It has been found that  Chern-Simons contribution to the bianchi identity
 has become cancelled from that of the scalar field part. But ``homogeneity and isotropy" consideration of present day cosmology is a consequence of the ``strong equivalence principle" and vice-versa.

\vspace{2mm}
   \noindent \textbf{\uppercase{pacs:}} 04.20.Fy, 04.20.Cv, 11.30.-j

\vspace{2mm}
   \noindent \textbf{\uppercase{key words:}}  Chern-Simons modified gravity, Torsion, Diffeomorphism invariance

 \end{abstract}

\section{Introduction}

Chern-Simons (CS) modified gravity  is a 4-dimensional deformation
of General Relativity (GR), postulated by Jackiw and
Pi\cite{Jac03}. This theory of  gravity is an extension of general
relativity by adding a parity violating Pontryagin density
${}^*RR$ coupled to a scalar field $\theta$ to Einstein-Hilbert
Lagrangian. This scalar field $\theta$ can be viewed as either a
prescribed background quantity or as an evolving dynamical field.
The  ${}^*RR$ is defined as the contraction of the Riemann tensor
with its dual and it is odd under a parity transformation, thus
potentially enhancing gravitational parity-breaking. The CS
correction introduces a means to enhance parity-violation through
a pure curvature term,  as opposed to through the matter sector,
as more commonly happens in GR. One of the important feature of CS
modified gravity is its emergence within predictive frameworks of
more fundamental theories. For example, the low energy limit of
string theory comprises general relativity with a parity violating
correction term, that is nothing but the Pontryagin density. This
term is crucial for cancelling gravitational anomaly in string
theory through Green-Schwartz mechanism\cite{Green01}. The
Pontryagin density, as an anomaly cancelling term, also arises in
particle physics and in the context of loop quantum
gravity\cite{Yun09}. Ref.\cite{Yun09} gives a  review on
Chern-Simons Modified General Relativity. In a  recent
approach the CS modified gravity reduces to topologically massive
gravity in three dimensions\cite{Ali10}.

It is now a well established fact that de Sitter group is the
correct underlying gauge group of gravity as unlike Poincare
group, it is a semisimple group yielding consistent field
equations. Thus de Sitter gauge theory comes up as the corrected
Poincare gauge theory\cite{Per88}. Recently a gravitational Lagrangian has been
proposed\cite{Mah02a}, where a
 Lorentz invariant part of the de Sitter Pontryagin density has been treated as
 the Einstein-Hilbert Lagrangian. In this formalism the role of torsion in the underlying manifold
is multiplicative
   rather than additive  and  the  Lagrangian  looks like
    $torsion \otimes curvature$.   This indicates that torsion is uniformly nonzero
   everywhere. In the geometrical sense, this implies that
   micro local space-time is such that at every point there is a
   direction vector (vortex line) attached to it. This effectively
   corresponds to the non commutative geometry having the manifold
   $M_{4}\times Z_{2}$, where the discrete space $Z_{2}$ is just
   not the two point space but appears as an attached direction vector\cite{Con94}. In this approach we consider only  a particular class of $U_{4}$ space,
    where  only the axial vector part of the  torsion is present everywhere in space time.\footnote{It is to be noted here that in the presence of spinorial matter only the axial vector
    part of the torsion couples to the spinor
    field\cite{Des94,Mie99}.}
   This may be compared with another approach\cite{Cantcheff} where the axial vector torsion is given by the derivative of a pseudoscalar field and then one gets a
    propagating torsion wave unlike in the standard Einstein-Cartan theory where torsion ceases to exist outside spinorial matter.
   This propagating torsion extends over whole spacetime. 
Some authors have  pointed that 
only scalar or psudoscalar torsion modes can be used to investigate the torsion cosmology\cite{Min06, Nes08}. Their conclusion emerges due to consideration of additive  torsion terms in the gravitational part of the Lagrangian. In particular the topological Nieh-Yan term, which is a psudoscalar density, is the term from the torsion sector which  plays an important role in the present formalism.
   Here  the additive torsion decouples from
   the theory but not the multiplicative one and this implies the non trivial omnipresence of  the
   axial  torsion. In this model of minimum extension of
   Einstein-Hilbert theory, we only take axial torsion as an extra degree of
freedom.   Considering torsion and torsion-less
  connection as independent fields\cite{Mah04}, it has been found that, $\kappa$  of Einstein-Hilbert Lagrangian appears as an integration constant and is linked with the topological Nieh-Yan density of $U_{4}$
  space. As a result, $\kappa$ has got its definite geometrical
  meaning in $U_{4}$ space in comparison to its standard meaning
  of being simply an ad hoc  constant.
  If we consider axial vector torsion together with a scalar field $\phi$ connected to a local  scale factor\cite{Mah05}, then the Euler-Lagrange equations
   also link, in laboratory scale, the mass of the scalar field with the Nieh-Yan density and, in cosmic scale of FRW-cosmology, they predict only three kinds of the phenomenological energy density representing mass, radiation and cosmological constant. In a recent paper\cite{Mah07}, it has been shown that this scalar field may
also be interpreted to be linked with the dark matter and dark radiation. Recently it has been shown
that\cite{Mah07a}, using field equations of all fields except the frame field, the starting
Lagrangian reduces to a generic $f(\mathcal{R})$ gravity Lagrangian which, for FRW
metric, gives standard FRW cosmology. But for non-FRW metric, in particular
of Ref.\cite{Lob07}, with some particular choice of the functions of the scalar field
$\phi$ one gets $f(\mathcal{R}) = f_0\mathcal{R}^{1+v^2{}_{tg}}$ , where $v_{tg}$ is the constant tangential velocity of
the stars and gas clouds in circular orbits in the outskirts of spiral galaxies.
  In this letter we are going to study the CS modification of  this formalism where the CS scalar $\theta$ has been considered to be a function of the scalar field $\phi$.
\section{Formulation}
The gravitational Lagrangian, with CS modification, may be defined
to be\cite{Yun09,Mah05,Mah07a}
\begin{eqnarray}\mbox{$\mathcal{L}$}_{G}+\mbox{$\mathcal{L}$}_{CS}&=& N\{ \mbox{$\mathcal{R}$}-u(\phi)\}+{}^*BB \nonumber\\&{}&\mp\frac{1}{2}d\phi\wedge{}^*d\phi -h(\phi)
\eta+\frac{1}{4}\theta{}(\phi)\hat{R}^{ab}\wedge\hat{R}_{ab},\label{eqn:abcd1}\end{eqnarray} where
  *   is Hodge duality operator,    $\mathcal {R}$$\eta=\frac{1}{2}\bar{R}^{ab}\wedge\eta_{ab}$, $B=B_a\wedge\bar{\nabla}e^a$, $\bar{R}^b{}_a=d\bar{\omega}^b{}_a+\bar{\omega}^b{}_c\wedge \bar{\omega}^c{}_a$, $\hat{R}^b{}_a=d\hat{\omega}^b{}_a+\hat{\omega}^b{}_c\wedge \hat{\omega}^c{}_a$,  $\bar{\omega}^{a}{}_{b}=\omega^{a}{}_{b}-T^{a}{}_{b}$, $\hat{\omega}^b{}_a= -e^{b\nu}
   e_{a\nu:\mu}dx^\mu$, $ T^a=\frac{1}{2!}e^{a\mu}T_{\mu\nu\alpha}dx^\nu\wedge dx^\alpha$, $T^{ab}=e^{a\mu}e^{b\nu}T_{\mu\nu\alpha}  dx^\alpha$, $T=\frac{1}{3!}T_{\mu\nu\alpha}dx^\mu\wedge dx^\nu\wedge dx^\alpha$, $N=dT$, $\eta_a=\frac{1}{3!}\epsilon_{abcd}e^b\wedge e^c\wedge e^d$ and $\eta_{ab}={}^*(e_a\wedge e_b)$. Here  $\bar{\nabla}$
represents exterior covariant differentiation with respect to the
connection one form $\bar{\omega}^{ab}$, $:$ represents tensorial
covariant differentiation, w.r.t. the Christoffel connection,
acting upon external indices and $B_{a}$ is a two form with one
internal index and of dimension $(length)^{-1}$ and $u(\phi)$,
  $h(\phi)$, $\theta(\phi)$ are unknown functions of $\phi$ whose
forms are to be determined subject to the geometric structure of
the manifold.

 Now we  write the total gravity Lagrangian in the presence of a spinorial matter field, given
by\begin{eqnarray}
\mbox{$\mathcal{L}$}_{tot.}&=&\mbox{$\mathcal{L}$}_{G}+\mbox{$\mathcal{L}$}_{CS}+\mbox{$\mathcal{L}$}_{D}, \label{eqn:apr}
\end{eqnarray} where\cite{Mah07}
\begin{eqnarray}
\mbox{$\mathcal{L}$}_{D}&=&[\frac{i}{2}\{\overline{\psi}{}^*\gamma\wedge D\psi+\overline{D\psi}\wedge{}^*\gamma\psi\}-\frac{g}{4}\overline{\psi}\gamma_5\gamma\psi\wedge T\nonumber\\&{}&+c_\psi\sqrt{{}^*dT} \overline{\psi}\psi\eta]\label{eqn:apr1}    \\
    \gamma_\mu &:=&\gamma_a e^a{}_\mu,\hspace{2mm}
    {}^*\gamma:=\gamma^a\eta_a,\hspace{2mm}
    D:=d+\Gamma\\
    \Gamma &:=&\frac{1}{4}\gamma^\mu D^{\{\}}\gamma_{\mu}=\frac{1}{4}\gamma^\mu \gamma_{\mu:\nu}dx^\nu\nonumber\\&=&-\frac{i}{4}\sigma_{ab}e^{a\mu}e^b{}_{\mu:\nu}dx^\nu
\end{eqnarray}
here $D^{\{\}}$, or $:$ in tensorial notation, is Riemannian torsion free covariant differentiation acting on external indices only; $\sigma^{ab}=\frac{i}{2}(\gamma^a\gamma^b-\gamma^b\gamma^a)$,  $\overline{\psi}=\psi^\dag\gamma^0$ and $g$, $c_\psi$ are both dimensionless coupling constants. Here $\psi$ and $\overline{\psi}$ have dimension $(length)^{-\frac{3}{2}}$ and conformal weight $-\frac{1}{2}$. It can be verified that under $SL(2,C)$ transformation on the spinor field and gamma matrices, given by,
\begin{eqnarray}    \psi\rightarrow\psi^\prime&=&S\psi,\hspace{2mm}\overline{\psi}\rightarrow\overline{\psi^\prime}=\overline{\psi}S^{-1}\nonumber\\\mbox{and}\hspace{2mm}\gamma\rightarrow\gamma^\prime&=&S\gamma S^{-1},
\end{eqnarray}where \mbox{$S=\exp(\frac{i}{4}\theta_{ab}\sigma^{ab})$,}
$\Gamma$ obeys the transformation property of a $SL(2,C)$ gauge connection, i.e.
\begin{eqnarray}
    \Gamma\rightarrow\Gamma^\prime&=&S(d+\Gamma)S^{-1}\\
    \mbox{s. t.}\hspace{2mm}D\gamma&:=&d\gamma+[\Gamma,\gamma]=0.\label{eqn:a47}
\end{eqnarray}

As in Ref.\cite{Mah05,Mah07},  by varying  the independent fields in the Lagrangian $\mbox{$\mathcal{L}$}_{tot.}$, we obtain the Euler-Lagrange equations and then after some simplification we get the following results
\begin{eqnarray}
&\bar{\nabla} e_{a}=0,\label{eqn:a470}\\&
{}^*N=\frac{1}{\kappa},\label{eqn:a471}
\end{eqnarray}
 i.e. $\bar{\nabla}$ is torsion free and $\kappa$ is an integration constant having  dimension of $(length)^{2}$.\footnote{In (\ref{eqn:abcd1}), $\bar{\nabla}$ represents a $SO(3,1)$ covariant derivative, it is only on-shell torsion-free through the field equation (\ref{eqn:a470}). The $SL(2,C)$ covariant derivative represented by the operator $D$ is torsion-free by definition, i.e. it is torsion-free both on on-shell and off-shell. Simultaneous and independent use of both $\bar{\nabla}$ and $D$ in the Lagrangian density (\ref{eqn:apr}) has been found to be advantageous in the approach of this article. This amounts to the emergence of the gravitational constant   $\kappa$ to be  only an on-shell  constant and this justifies the need for the introduction of the Lagrangian multiplier $B_a$ which appears twice in the Lagrangian density (\ref{eqn:abcd1}) such that $\bar{\omega}^a{}_b$ and $e^a$  become independent fields.}
\begin{eqnarray}   & m_\psi=c_\psi\sqrt{{}^*dT}=\frac{c_\psi}{\sqrt{\kappa}},
\\&i{}^*\gamma\wedge D\psi-\frac{g}{4}\gamma_5\gamma\wedge T\psi+m_\psi\psi\eta=0,&\nonumber\\&
i\overline{D\psi}\wedge{}^*\gamma-\frac{g}{4}\overline{\psi}\gamma_5\gamma\wedge T+m_\psi\overline{\psi}\eta=0.&
\end{eqnarray}

\begin{eqnarray}
(G^b{}_a+C^b{}_a)\eta&=&-\kappa[ \frac{i}{8}\{\overline{\psi}(\gamma^b D_a+\gamma_a D^b)\psi-(\overline{D_a\psi}\gamma^b+\\&{}&\overline{D^b\psi}\gamma_a)\psi\}\eta-\frac{g}{16}\overline{\psi}\gamma_5(\gamma_a {}^*T^b+\gamma^b {}^*T_a)\psi\eta\\&{}&\pm\frac{1}{2}\partial_a\phi\partial^b\phi\eta+\frac{1}{2}h\eta\delta^b{}_a]\\&=&-\kappa[T^b{}_a(\psi)+T^b{}_a(\phi)]\eta\hspace{2 mm}\mbox{say},\\0&=&[\frac{1}{2}  \overline{\nabla}_\nu\overline{\psi}\{\frac{\sigma^b{}_{a}}{2}, \gamma^\nu\} \psi+\frac{i}{2}\{\overline{\psi}(\gamma^b D_a-\gamma_a D^b)\psi\nonumber\\&{}&-(\overline{D_a\psi}\gamma^b-\overline{D^b\psi}\gamma_a)\psi\} \\&{}&-\frac{g}{4}\overline{\psi}\gamma_5(\gamma_a {}^*T^b-\gamma^b {}^*T_a)\psi]\eta,\\C^{ab}&=&\kappa[\overline{\nabla}_c\theta\epsilon^{cde(a}\bar{\nabla}_e\mathcal{R}^{b)}{}_d+\overline{\nabla}_c\overline{\nabla}_d\theta{}^*\bar{R}^{d(ab)c}],\\T^b{}_a(\psi)&=& \frac{i}{8}\{\overline{\psi}(\gamma^b D_a+\gamma_a D^b)\psi-(\overline{D_a\psi}\gamma^b+\\&{}&\overline{D^b\psi}\gamma_a)\psi\}-\frac{g}{16}\overline{\psi}\gamma_5(\gamma_a {}^*T^b+\gamma^b {}^*T_a)\psi,\\T^b{}_a(\phi)&=&\pm\frac{1}{2}\partial_a\phi\partial^b\phi+\frac{1}{2}h\delta^b{}_a,
\end{eqnarray}
\begin{eqnarray} \kappa d[\frac{g}{4}{}^*(\overline{\psi}\gamma_5\gamma\psi\wedge T)+\Sigma]=-\frac{g}{4}\overline{\psi}\gamma_5\gamma\psi,\\\Sigma=\mp \frac{1}{2}{}^*(d\phi\wedge{}^*d\phi)+2h-\frac{1}{\kappa}u \\\{\frac{1}{\kappa}u^\prime(\phi) -h^\prime(\phi)\}
\eta+\frac{1}{4}\theta^\prime(\phi)\hat{R}^{ab}\wedge\hat{R}_{ab}=\mp
d{}^*d\phi.\label{eqn:25}
\end{eqnarray}
Now we see that\cite{Jac03,Yun09}
\begin{eqnarray}
\bar{\nabla}_bC^b{}_a&=&-\frac{1}{8}\kappa\partial_a\theta{}^*RR\\
\bar{\nabla}_bT^b{}_a(\phi)&=&\bar{\nabla}_b[\pm\frac{1}{2}\partial_a\phi\partial^b\phi+\frac{1}{2}h\delta^b{}_a]=\frac{1}{8}\partial_a\theta{}^*RR+\frac{1}{2}\partial_a\Sigma
\end{eqnarray}Therefore Bianchi identity of $G^b{}_a$ implies
\begin{eqnarray}
\bar{\nabla}_bT^b{}_a(\psi)=-\frac{1}{2}\partial_a\Sigma\label{eqn:28}
\end{eqnarray}

Earlier\cite{Mah05,Mah07,Mah08}, we have seen that $\Sigma=
constant$ gives us standard isotropic and homogeneous
FRW-universe at cosmic scale and from the last equation we see
that this is the case of strong equivalence principle. Therefore
we can state that \textbf{Isotropy \& Homogeneity
$\Leftrightarrow$ Diffeomorphism Invariance}. Moreover, for
$\Sigma= constant$,  we can define conserved axial
current\cite{Mah08} by
 \begin{eqnarray}
 J\equiv\kappa {}^*(\overline{\psi}\gamma_5\gamma\psi\wedge T)T
\end{eqnarray}

The general theory of relativity is a diffeomorphism invariant
theory where energy momentum tensor is covariantly conserved and
there exists a killing field that generates an isometry of the
spacetime. Now the vital question is which solution of Einstein's
equation corresponds to our universe or at least an idealized
model that approximates our universe. We know that the structure
of the universe as predicted by GR based on simple cosmological
principle of homogeneity and isotropy approximation in the  large
scale structure of our universe. So diffeomorphism invariance of
Einstein's equation and simple cosmological principle are closely
related. In our present formalism of minimal extension of GR we
see that diffeomorphism invariance is still maintained if and only
if the universe is isotropic and homogeneous. The equation
(\ref{eqn:28}) is valid with or without CS term of the action. So
this conclusion remains valid even in the presence of CS modified
extended GR.  When we consider other form of cosmic energy
density, may be in the early universe, we have to adopt a non-FRW
geometry where we may have to forgo the isotropy and homogeneity
of the universe\cite{Mah07}, then the above result shows that
conservation of energy-momentum tensor of baryonic matter is
violated but the total energy-momentum tensor of both baryonic and
dark matter/radiation is always conserved.

\section{Discussion}
In the standard formulation of Chern-Simons modified gravity where
$\theta$ is taken as external variable, the presence of Cotton
tensor $C^b{}_a$ in the modified Einstein's equation violates the
diffeomorphism invariance because the non vanishing divergence of
cotton tensor is proportional to $*RR$, the Pontryagin density. To
maintain the diffeomorphism invariance the consistency of dynamics
forces $*RR$ to vanish, so the consistency condition suppressed
the symmetry breaking CS term in the action, even though its
variation results in the modified equation of
motion\cite{Jackiw02}. When $\theta$ is taken as local dynamical
variable then this constraint is replaced by evolution equation
of $\theta$ which can be viewed as Klein-Gordon equation in the
presence of a potential and Pontryagin density as source
term\cite{Yun09}. Then from modified Einstein's equation it can be
shown that strong equivalence principle is satisfied provided
$\theta$ satisfies its evolution equation. So there is no need to
impose Pontryagin constraint to maintain Lorentz symmetry and
thereby strong equivalence principle.

In the first order formalism of CS modified
gravity\cite{Cantcheff}, curvature tensor is taken as $SO(3,1)$
field strength  such that the gravitational part gives the
standard first order Einstein-Hilbert action but the CS part
${}^*RR$, being not torsion free,  serves as the effect of the CS
deformation on the space-time geometry  through an effective
contribution of torsion which depends on the external field
$\theta$\cite{Cantcheff}. In the absence of CS term the action
reduces to standard Einstein-Hilbert action. But in our model of
minimal extension of Einstein-Hilbert action a multiplicative
torsion is already present in the lagrangian which looks like
$torsion \otimes curvature$. The additive torsion may decouples
from the theory but not the multiplicative one. This indicates
that torsion is uniformly nonzero everywhere. This torsion is not
present in the curvature tensor and in the CS term. Here we take
the CS term as $\theta{}(\phi)\hat{R}^{ab}\wedge\hat{R}_{ab}$,
where $\hat{R}^{ab}$ is the curvature 2-form from the metric only.

In our formulation of minimal extension of GR the evolution
equation for $\phi$ is given by equation (\ref{eqn:25}). Here we
see that strong equivalence principle is not automatically satisfied but it
depends upon the isotropy and homogeneity of spacetime
(\ref{eqn:28}). This is because here we take CS modification of
minimal extended gravity, $\phi$ is taken as dark matter field and
$\theta$ is taken as a function of $\phi$. As a consequence we see
that energy momentum conservation of ordinary (baryonic) matter
depends upon the constancy of $\Sigma$ (or isotropy and
homogeneity of spacetime). And therefore strong equivalence
principle of ordinary baryonic matter depends entirely upon the
evolution of dark matter to form the FRW-geometry of the present
universe in cosmic scale.

\end{document}